\documentstyle[titlepage,12pt]{article}

\newcommand{\beq}{\begin{equation}}
\newcommand{\eeq}{\end{equation}} 
\newcommand{\eq}[1]{eq.(\ref{#1})}

\title{\bf The Lamb Shift of Excited $S$-Levels in Hydrogen and Deuterium Atoms}

\author{Savely~G.~Karshenboim\thanks{ E-mail: sgk@onti.vniim.spb.su;
karshenboim@phim.niif.spb.su}${}^,$\thanks{The summer address: 
Max-Planck-Institut f\"ur
Quantenoptik, 85748 Garching, Germany. E-mail: krp1@mpq.mpg.de}\\ 
D.~I.~Mendeleev Institute
for Metrology (VNIIM)\\ 
St.~Petersburg 198005, Russia}
\date{}

\begin{document}

\maketitle

\begin{abstract}

A specific combination of $s$-state Lamb shift $
\Delta E_L(1s_{1/2}) - n^3 \Delta E_L(ns_{1/2})$ is considered. Its value 
in calculated both in the hydrogen and
deuterium atoms for $n$ up to 12. The result inludes all correction which 
can contribute $1\, kHz$ and
particulary: one-loop self energy and vacuum polarization, two-loop contribution. Nuclear finite-size
corrections for the isotopic difference of the combination are also evaluated.   

\end{abstract}

\section{Introduction}

Recently new experimental results on the { hydrogen} and { deuterium Lamb
shift} have been obtained \cite{Sch93}--\cite{Bou95} and some even
higher-precision experiments are going to be completed \cite{PrH,PrB}. The each
measurement data include a combination of the Lamb shift of several
$ns_{1/2}$  levels ($n=1\div 12)$. To obtain correlations between theoretical calculations of the Lamb shift
of $s$-states with  { different} values of principal quantum number {$ n$} or
their combinations, the specific difference of Lamb shifts

\beq  \label{del} 
\Delta(n) = \Delta E_L(1s_{1/2}) -
  n^3 \Delta E_L(ns_{1/2})
\eeq

\noindent is considered in this work.

The values $\Delta(n)$ have to be used to find { self-consistent} values of the
{ ground state} Lamb shift and the { Rydberg constant} \cite{JETP94,NP951}.


\section{General expression and definitions}


The Lamb shift is defined  here as a shift from the value of level energy

\beq \label{definitio}
  E(nl_j)=
  m_r \left[ f(nj)-1 \right]
  -\frac{m_r^2}{2(M+m)} \left[ f(nj)-1\right ]^2,
\eeq

\noindent where

\[
  f(nj) =
  \left( 1+\frac{(Z\alpha)^2}
  { \left[ n - j - 1/2 + \sqrt{(j+1/2)^2-(Z\alpha)^2} \right]^2 }
  \right)^{-1/2}
\]

\noindent
is the dimensionless Dirac energy with infinite nuclear mass, $m_r$ is the reduced mass,
$Z$ is the nuclear charge in units of the proton one, and relativistic units in which
$ \hbar  = c = 1$ and $ \alpha=e^2 $ are used.

The general expression for \eq{del} has the form \cite{JETP94,NP951}

\beq   \label{LS}
\Delta(n) =
\frac{\alpha(Z\alpha)^4}{\pi}\frac{m_r^3}{m^2}\times
\Bigg\{- \frac{4}{3}\log{\frac{k_0(1s)}{k_0(ns)}}
\left(1+Z\frac{m}{M}\right)^2 +(Z\alpha)^2\times
\]

\[
\Bigg[\left(4\big(\log{n}-
\psi(n+1)+\psi(2)\big)-
\frac{77(n^2-1)}{45 n^2}\right)
\log{\frac{1}{(Z\alpha)^2}}+
A^{VP}_{60}(n)+G^{SE}_{n}(Z\alpha)\Bigg]
\]

\[
-\frac{14}{3}\frac{Zm}{M} \left(\psi(n+1)-\psi(2)
-\log{n}+\frac{n-1}{2n}\right)\Bigg\}+
\frac{\alpha^2 (Z\alpha)^6m}{\pi^2}
\log^2{\frac{1}{(Z\alpha)^2}}\,B_{62}
,\eeq

\noindent where $\psi(z) = (d/dz)\log\Gamma(z)$.


\medskip

\noindent Here\\

{\Large$\bullet$} $\log{k_0(ns)}$ is the Bethe logarithm, its value can be found in Refs.
\cite{DSw90,Kla73};\\

\smallskip {\Large$\bullet$} a logarithmic coefficient in the term
$\alpha(Z\alpha)^6\log(Z\alpha)\, m$ was obtained in Ref. \cite{EY};\\

\smallskip {\Large$\bullet$}
$G^{SE}_n(Z\alpha)$ is { one}-loop { self-energy} correction in the order of 
$\alpha(Z\alpha)^6 m$ and  higher;\\

\smallskip
{\Large$\bullet$}
$A^{VP}_{60}(n)$ is the $\alpha(Z\alpha)^6 m$-contribution of the { vacuum polarization};\\

\smallskip
{\Large$\bullet$}
$B_{62}$ is the { leading} logarithmic { two}-loop correction coefficient;\\

\smallskip
{\Large$\bullet$} the three-loop term of the order $\alpha^3(Z\alpha)^4m$  in \eq{LS} is known and
equal to zero;\\

\smallskip
{\Large$\bullet$} recoil corrections in orders  $(Z\alpha)^4 m^2/M$  and $(Z\alpha)^6
m^3/M^2$ are equal to zero;\\

\smallskip
{\Large$\bullet$} the nuclear finite-size contributions for the difference
(\ref{LS}) are small and not included.\\

The Bethe logarithm is presented in  Table 1. 
The one- and two-loop contributions and recoil corrections are discussed 
below. The nuclear charge
distribution correction is considered in appendix B.


\section{One-loop self-energy contribution}

\subsection{Extrapolation over $Z$}


The values of $G^{SE}_n(Z\alpha)$ with $n=2\div5$ for the { hydrogen} atom can be derived from the
numerical data given in Refs. \cite{Moh921,Moh922} after { extrapolation} to $Z=1$, with including
for the difference  (\ref{del}) { only two terms} \cite{JETP94,NP951}

\beq  \label{fit}
G^{SE}_n(Z\alpha) = A^{SE}_{60}(n)+ (Z\alpha) A^{SE}_{70}(n)
,\eeq

\noindent
because the logarithmic coefficient 
in the order $\alpha(Z\alpha)^7 m$
has the form \cite{JETP94}

\[
A^{SE}_{71}(nl_j) =
\pi \left(\frac{139}{64}-\log{2}
\right) \delta_{l0}
\]

\noindent
and does not contribute to the difference $\Delta(n)$. In case of $n\!=\!2$ a result from Ref.
\cite{Pac93} for $A^{SE}_{60}(2)$ has been also used.

The input data and the results of the extrapolation are given in Table 2. 

The  quoted { uncertainties} arise from { statistical} errors of the numerical integrations of
Refs. \cite{Moh921,Moh922} and from  estimates of the { systematic} error of the fit of \eq{fit}.


\subsection{Extrapolation over $n$}


The extrapolated value of $G^{SE}_n(\alpha)$ can be rewritten in the form

\beq
G^{SE}_n(\alpha) =
\frac{n-1}{n^2}\,\widetilde{G}_n
\;
,\eeq

\noindent
which is more convenient for the { extrapolation} over {$ n$}. This
transformation is needed because the $n$-dependence of $G^{SE}_n(\alpha)$ is not
well-behaved. That is easy to see from the trivial result

\[G^{SE}_1(\alpha)=0.\] 

\noindent
After extracting the factor $(n\!-\!1)$ { explicitly} the dependence of 
$\widetilde{G}_n$ over $n$ { became} quite  well-behaved, as well as in case of other coeffitients of
\eq{LS}.

Our results for $n=6\div 12$ are presented in Table 3. 



\section{One-loop vacuum polarization}


As it was demonstrated in Ref. \cite{IK96NP}, for a calculation of the contribution to the 
{ difference}
$\Delta(n)$  it is sufficient to use the approximate wave function

\beq \label{apr1}
  \psi = \left(
    \begin{array}{c}
       1 \\
       \frac{ {\bf p } \sigma  }
            {2m  }
    \end{array}
  \right)\,
 \varphi,
\eeq

\noindent
where $\varphi$ is the Schr\"odinger wave function, and the effective local potential

\beq \label{apr2}
  V_{VP}( \hbox{\bf r} ) = \frac{\alpha (Z\alpha)}{m^2}
    \left(
      -\frac{4}{15} \delta( \hbox{\bf r} )
      -\frac{1}{35} \frac{\nabla^2}{m^2} \delta( \hbox{\bf r} )
    \right).
\eeq

The { difference} coefficient of the vacuum polarization contribution is found to be \cite{IK96NP}

\beq \label{avp60}
A^{VP}_{60}(n) =  \frac{4}{15}
\left\{
\log{n}-
\psi(n+1)+\psi(2)
+\frac{n^2-1}{28n^2}
+\frac{2(n-1)}{n^2}
\right\}
.\
\eeq

Results for $n=2$ and $n=4$ are in agreement with those from Refs. \cite{v2} and \cite{v4}. The
approximation of eqs. (\ref{apr1}, \ref{apr2}) leads to the same result as in work \cite{MNF}
for $np_j$-states.

\section{Two-loop corrections}


The leading two-loop contribution has order $\alpha^2 (Z\alpha)^6m\log^2(Z\alpha)$. It originates from
two-loop self energy of an electron in the Coulomb field 
\cite{JETP94,NP951}, a general expression of which  has the form \cite{Mil}

\beq   \label{sig}
\delta E^{(2)}_L(nl_j)=\langle nl_j\vert
\Sigma_1(E_{nl_j})\overline{G}_C(E_{nl_j})
\Sigma_1(E_{nl_j})\vert nl_j\rangle
\]

\[
+\langle nl_j\vert \Sigma_2(E_{nl_j})\vert nl_j\rangle
 +
\langle nl_j\vert \Sigma_1(E_{nl_j})\vert nl_j\rangle
\langle nl_j\vert \frac{\partial\Sigma_1(E)}{\partial E}\vert
nl_j\rangle \Big\vert_{E=E_{nl_j}},
\eeq

\noindent
where $\Sigma_r(E)$ is the $r$-loop one-particle-irreducible self-energy
operator of an electron in the Coulomb field and $\overline{G}_C(E)$ is the
reduced Coulomb Green«s function.

In the { Yennie gauge} \cite{Abr,FY}, in which the photon propagator in
momentum space has the form

\[
D^{Y}_{\mu\nu} (k) = \frac{1}{k^2}
\left(
g_{\mu\nu}+2\frac{k_\mu k_\nu}{k^2}
\right),
\]

\noindent
the result arises from only the first item of \eq{sig} and the ladder part of the 
second one. Contributions of separated diagrams  to different energy
levels are presented in  Table 4.

The contribution was found in Refs.
\cite{JETP94,NP952,JP96,JETP96} and finally it has the form

\beq        \label{FinLs}
B_{62} =
\frac{16}{9} \left( \log{n} - \psi(n)+\psi(1) -\frac{n-1}{n} +\frac{n^2-1}{4n^2} \right)
.
\eeq

\noindent The uncertainty of two-loop correction is due to terms beyond $\log^2{Z\alpha}$ and it is
estimated as  { half the contribution} of the leading logarithmic term.

The same diagrams contribute  in the order $\alpha^2(Z\alpha)^6
m \log{Z\alpha}$ also to the { decay widths} of $ns$- and 
$np$-levels (see for details Ref. \cite{Dipol}), which can be presented as the 
imaginary part of the self-energy of an electron.
This fact is useful for { checking} \cite{JETP96}. An expecit expression of
$\overline{G}_C(0,{\bf r};E_{nl})$ found in Ref. \cite{Dipol} has also been used for that.

\section{Recoil corrections}

The evaluation of { pure recoil} contributions has recently been  { completed} for $s$-levels
in the orders  $(Z\alpha)^6 m^2/M$ \cite{PG,Yel} and $(Z\alpha)^4 m^3/M^2$  \cite{PK95}. 

The results for the term of order $(Z\alpha)^6 m^2/M$

\beq                  \label{PacZ6}
\delta E (ns) = \frac{(Z\alpha)^6m^2}{n^3M}
\left(4\ln{2}-\frac{7}{2}\right),~~~~~\cite{PG},
\eeq

\noindent and

\beq                  \label{YelZ6}
\delta E (ns) = \frac{(Z\alpha)^6m^2}{n^3M}
\left(4\ln{2}-\frac{5}{2}\right),~~~~~\cite{Yel},
\eeq

\noindent 
are in {
disagreement}. In our definitions (\ref{definitio}) { both} of them do { not} contribute to the
difference $\Delta(n)$. Numerical results for recoil correction in the $1s$- and $2s$-states obtained
in Ref. \cite{Art95} without any expanssion over $(Z\alpha)$ confirm the scaling factor $1/n^3$
(see table 5) and they are in fair agreement with \eq{PacZ6}. It should be also menshioned that results of
Yelkhovsky \cite{Yel} and Pachucki \cite{PacZ6l} for higher-$l$ levels 
and in agreement one with the other, and with
the analytic result for $p$-state of Ref. \cite{Gol} and the numerical ones for $2p_{1/2}$ and 
$2p_{3/2}$ of Refs. \cite{Art95} and
\cite{ASY2}, respectively.

The correction in order $(Z\alpha)^4 m^3/M^2$ contributes only for 
nuclear spin $I=0$ or 1, but not for $I=1/2$
\cite{PK95}. The result is 

\beq
\delta E (I = 0 ,1 ) = - \frac{1}{2} \frac{m^3}{M^2}
\frac{(Z\alpha)^4}{n^3}\delta_{l0}.
\eeq

All recoil corrections in this section are equal to zero for the difference (\ref{del}), but no
proof on these cancelations is not known without direct calculations.

\section{Results}


The differences $\Delta(n)$ and the contributions to it for the lowest $ns$-levels of hydrogen
and deuterium are presented in Table 6.

The { uncertainty} of the theoretical expression is a rms sum of the
uncalculated terms of orders $\alpha^2(Z\alpha)^6m\log(Z\alpha)$ and
$\alpha^2(Z\alpha)^6m$, which are estimated as { half the contribution} of the
leading logarithmic term and the uncertainty of the { extrapolation} for the one-loop self-energy
contribution. The uncertainty of the isotopic difference

\[
\Delta^{Iso}(n)=\Delta^{Deu}(n) - \Delta^{Hyd}(n)
\]

\noindent
is neglegible and the nuclear finite-size corrections could be important for it. 
They are considered in appendix B and
included in the results in Table 6.

\section*{ ACKNOWLEDGEMENTS}

The author would like to thank T. W. H\"ansch, A.~Nekipelov, K.~Pachucki, 
T. Udem, V. Shabaev, M.~Weitz and A. Yelkhovsky
for stimulating discussions and the authors of Refs. 
\cite{Wei95,Ber95,Bou95,PrH,PrB,v4,PrP,Yel,Art95,ASY2} for informing him on
their results prior the publication. The final part of this work was done 
durinig the author«s summer stay at the Max-Planck-Institut f\"ur Quantenoptik and he 
is very grateful them for hospitality. 

This work was supported in part by grant \# 95-02-03977 of the Russian Foundation for
Basic Research.  The part of the work was done by collaboration with  
V.~Ivanov and presented at the CPEM96, and  the author acknoledges him for the cooperation and
their organizing commitee for support of the poster. The author thanks the 28$^{th}$EGAS Organizing Commitee for support of his
participation a the conference, where this work was also 
presented.

\newpage

\appendix

\section{Definitions}

The definition of the Lamb shift which is used in this work
is due to a shift from the energy level given in \eq{definitio}. It is 
denoted in this Appendix as $E^{I}(nj)$.

The expansion of $E(nj)$ has the form 

\[
E = m\sum_{0}^{\infty}{\left(\frac{m}{M}\right)^a(Z\alpha)^{b}\,E_{ab}}.
\]

An other definition which is also often used  
include only all $E_{0b}$ (i. e. the exact Dirac energy with the infinite nuclear mass) terms and $E_{a2}$
(the Schr\"odinger energy with reduced mass)  and a relativistic recoil term $E_{14}$:

\beq  \label{def2}
E^{II}(nj) = m\,f(nj)+\big[m_R-m\big]\,\left(-\frac{(Z\alpha)^2}{2n^2}\right)
+\frac{(Z\alpha)^4}{2n^3}\,\frac{m}{M}\,\left[\frac{1}{j+1/2}-\frac{1}{n}\right]
.
\eeq

Hence the main exluded items in value $E^{II}$ are

\[
E_{16} = -\frac{1}{2}\left[-\frac{2}{n^3(2j+1)^3}
-\frac{3}{n^4(2j+1)^2}+\frac{4}{n^5(2j+1)}-\frac{1}{n^6}\right]
\]

\noindent and

\[
E_{24} = \frac{1}{2n^3}\left[-\frac{1}{j+1/2}+\frac{3}{2n}\right].
\]

Advantages of the first definition of \eq{definitio} are: extra $E_{16}$ terms do
not contribute to $\Delta(n)$ and to the fine structure.Also extra $E_{24}$ terms
do not contribute to $E(ns)$ (for nuclear spin $=$ 1/2) and to the difference 
$\Delta(n)$ for any nuclear spin.

Difference between definitions of  \eq{definitio} and  \eq{def2}  in case 
of the hydrogen atom leads to the shifts  presented in
Table 7.

\section{Nuclear charge distribution correction}

Using the nonrelativistic approximation (the same as in case of vacuum polarisation) the nuclear charge
distribution correction to the difference can easy be found to be (cf. \cite{IK96NP})

\[
\delta \Delta (n) = {\cal E}(1s) \times (Z\alpha)^2
\left[ \psi(n+1)-\psi(2) - \log{n} - \frac{(n-1)(n+9)}{4n^2} \right]
,\]

\noindent where ${\cal E}(1s)$ is the well-known nuclear charge
distribution correction to the energy of the $1s$ state

\[
{\cal E}(1s) = \frac{2}{3}(Z\alpha)^4m_R^3 \langle r^2 \rangle.
\]

One can see that the correction to difference is small and can be neglected both for hydrogen and deuterium
atom, but for isotopic difference. They are slightly $n$-dependent 
(cf. term  $\Delta^{VP}(n)$ in table 6) and
the contribution to $\Delta(n)$ $(n=2-8)$ can be estimated as $-0.049(9)\,kHz$
for hydrogen,
$-0.318(33)\,kHz$ for deuterium and as $-0.269(32)\,kHz$ for the isotopic
difference. The uncertainty takes into acount both: $n$-dependence and the
disagreements in proton \cite{Han,Sim,Mer}  and deuteron \cite{Kla86,Fri93} radii
measurements.

\newpage

\begin{table}
\begin{center}
\vspace*{5mm}
\begin{tabular}{||c|c|c||}
\hline
\hline
&&\\[-1ex]
Level & $\log{k_0(nl)}$ &  Ref. \\[1ex]  
\hline
&&\\[-1ex]
$1s$  & ~2.9841285558~  & \cite{Kla73}  \\ [1ex] 
$2s$  & ~2.8117698931~  & \cite{Kla73}  \\ [1ex] 
$3s$  & ~2.7676636125~  & \cite{Kla73}  \\ [1ex] 
$4s$  & ~2.7498118405~  & \cite{Kla73}  \\ [1ex] 
$5s$  & ~2.7408237279~  & \cite{DSw90}  \\ [1ex] 
$6s$  & ~2.7356642069~  & \cite{DSw90}  \\ [1ex]  
$7s$  & ~2.7324291292~  & \cite{DSw90}  \\ [1ex] 
$8s$  & ~2.7302672607~  & \cite{DSw90}  \\ [1ex] 
$9s$  & ~2.7287511660~  & \cite{DSw90}  \\ [1ex] 
$10s$ & ~2.7276469387~  & \cite{DSw90}  \\ [1ex] 
$11s$ & ~2.7268177825~  & \cite{DSw90}  \\ [1ex] 
$12s$ & ~2.7261793406~  & \cite{DSw90}  \\ [1ex] 
\hline
\hline
\end{tabular}
\end{center}
\caption{Bethe logarithms.       }
\end{table}

\newpage

\begin{table}
\vskip 1cm
\begin{center}
\begin{tabular}{||r|l|l|l|l||}
\hline
\hline
\multicolumn{5}{||c||}{}\\[-1ex]
 \multicolumn{5}{||c||}{ Input data for the extrapolation over $Z$}\\[1ex]
\hline
&&&&\\[-1ex]
$Z$ & $G^{SE}_2(Z\alpha)$ & $G^{SE}_3(Z\alpha)$ & $G^{SE}_4(Z\alpha)$ & $G^{SE}_5(Z\alpha)$ \\[1ex]
\hline 
&&&&\\[-1ex]
0 & 0.912${}^a$  &   &  & \\[1ex]
\hline 
&&&&\\[-1ex]
5 & 0.860 (150)${}^b$ &           &            &             \\[1ex]
10 & 0.707 (11)${}^b$ & 0.601 (38)${}^{b,c}$ & 0.492 (75)${}^{b,c}$  &  0.412
(113)${}^{b,c}$ \\[1ex] 
15 & 0.619 (1)$^b$  &           &            &            \\[1ex] 
20 &
0.530${}^b$      & 0.436 (5)${}^{b,c}$  & 0.349 (5)${}^{b,c}$   &  0.286
(9)${}^{b,c}$   \\[1ex] 
25 & 0.445${}^b$      &           &           
&             \\[1ex] 
30 & 0.362${}^b$      & 0.284 (2)${}^{b,c}$  & 0.220
(2)${}^{b,c}$   &  0.175 (2)${}^{b,c}$   \\[1ex] 
\hline
\multicolumn{5}{||c||}{}\\[-1ex]
\multicolumn{5}{||c||}{ Results of the extrapolation over $Z$}\\[1ex]
\hline 
&&&&\\[-1ex]
1 & 0.89 (2) & 0.75 (17) & 0.62 (21)  &  0.53 (27) \\[1ex]
\hline\hline 
\end{tabular}
\end{center}
\caption{Extrapolation over $Z$. ${}^{a}$ -- from Ref. [15]; 
${}^{b}$ -- from Ref.  [14];
${}^{c}$ -- from Refs.  [13,14]
.}
\vskip 1cm
\end{table}

\begin{table}
\vskip 0.5cm
\begin{center}
\begin{tabular}{||c|c|c||}
\hline
\hline
&&\\[-1ex]
$n$  & $G^{SE}_n(\alpha)$  & $\widetilde{G}_n$\\[1ex]
\hline
&&\\[-1ex]
2 & 0.89(02) & 3.56(09) \\[1ex]
3 & 0.75(17) & 3.38(78) \\[1ex]
4 & 0.62(21) & 3.31(114) \\[1ex]
5 & 0.53(27) & 3.29(167) \\[1ex]
\hline
&&\\[-1ex]
6 & 0.46(28) & 3.30(200) \\[1ex]
7 & 0.40(31) & 3.30(250)\\[1ex]
8 & 0.36(33) & 3.30(300)\\[1ex]
9 & 0.33(35) & 3.30(350) \\[1ex]
10 & 0.30(36) & 3.30(400)\\[1ex]
11 & 0.27(37) & 3.30(450)\\[1ex]
12 & 0.25(38) & 3.30(500) \\[1ex]
\hline\hline
\end{tabular}
\end{center}
\caption{Extrapolation over $n$.}
\vskip 1cm
\end{table}

\pagebreak

{%
\small
\begin{table}
\footnotesize
\begin{center}
\vspace*{5mm}
\begin{tabular}{||c|c|c||}
\hline
\hline
&&\\
& $<\Sigma_1G_C\Sigma_1>$ &$<\Sigma_2^{lad}>$  \\
&&\\
\hline
\hline
&&\\
$ E(ns)$&$
-\frac{8}{27\pi^2}\frac{1}{n^3}m \log^3{\frac{1}{(Z\alpha)^2}}$,~~$^a$
&$0$\\
&&\\
\hline
&&\\
$\Delta (n)$&$
\frac{16}{9\pi^2}
\left(\log{n} -\psi(n+1)+\psi(2)- \frac{n-1}{n}\right)
 \log^2{\frac{1}{(Z\alpha)^2}}
$,~$^{b,c}$
&$
\frac{4}{9\pi^2}\frac{n^2-1}{n^5}
\log^2{\frac{1}{(Z\alpha)^2}}$,~~$^d$ \\
&&\\
\hline
&&\\
$E(np)$&$0$&$
\frac{4}{27\pi^2}\frac{n^2-1}{n^5} 
\log^2{\frac{1}{(Z\alpha)^2}}$,~~$^d$\\
&&\\
\hline
&&\\
$\Gamma (2p) $ & 0
&$- 
\frac{2^{10}}{3^9\pi} \ln{\frac{1}{(Z\alpha)^2}}
\left(- 2\ln{\frac{4}{3}} + \frac{61}{24} \right)$,~$^{b,e,f,g,h}$\\
&&\\
\hline
&&\\
$\Gamma (3s)$ &$-  
\frac{2^{10}}{ 5^9\pi} \ln{\frac{1}{(Z\alpha)^2}}
\left(2\ln{\frac{5}{4}} + \frac{387}{40} \right)$,~~$^{f,h}$&0\\
&&\\
\hline
\hline
\end{tabular}
\end{center}
\caption{Leading logarithmic contributions of two-loop self-energy in the Yennie gauge to different
energy levels and decay widths ($\Gamma$). Results are done in units of $\alpha^2(Z\alpha)^6m $.
$^a$ -- Ref.  [23]; 
$^b$ -- Ref.  [8]; 
$^c$ -- Ref.  [24]; 
$^d$ -- Refs.  [25,26]; 
$^e$ -- Ref.  [27] 
(numerically); 
$^f$ -- Ref.  [28] 
(numerically); 
$^g$ -- Ref.  [29] 
(analytically and numerically);
$^h$ -- Ref. [30] 
(analytically).
}
\vspace*{5mm}
\end{table}
}

\pagebreak

\begin{table}

\vskip 0.5cm
\begin{center}
\begin{tabular}{||c|c|c|c||}
\hline
\hline
&&&\\[-1ex]
Ref. &$\delta E(1s)$&$\delta E(2s)$&$\delta E(1s)/8\delta E(2s)$\\[-1ex]
&&&\\[1ex]
\hline
&&&\\[-1ex]
\cite{PG}&-7.4&-0.93&1\\[1ex]
\cite{Yel}&2.8&0.35&1\\[1ex]
\cite{Art95}&-7.1(9)&-0.90(6)&0.99(13)\\[1ex]
\hline\hline
\end{tabular}
\end{center}
\caption{Recoil correction in  order  $(Z\alpha)^6 m^2/M$ in the hydrogen atom in kHz. Numerical results of
Ref. [34] 
include also higher order in $(Z\alpha)$ corrections.
}
\vskip 1cm
\end{table}

\begin{table}
\vskip 1cm

\begin{center}
\begin{tabular}{||c|c|c|c|c|c|c||}
\hline
\hline
&&&&&&\\[-1ex]
$n$ & $\Delta^{G}(n)$ & $\Delta^{VP}(n)$ & $\Delta^{II}(n)$
 & $\Delta^{Hyd}(n)$ & $\Delta^{Deu}(n)$  &$\Delta^{Iso}(n)$\\[1ex]
\hline 
&&&&&&\\[-1ex]
2 & 39 (1)   &  8 & -11(5) & -187232(5)  & -187225(5) & 7.3 \\[1ex]
3 & 33 (7)   &  8 & -15(7) & -235079(10) & -235073(10)& 5.9 \\[1ex]
4 & 27 (9)   &  9 & -17(8) & -254428(12) & -254423(12)& 4.7 \\[1ex]
5 & 22 (11)  &  8 & -18(9) & -264162(15) & -264158(15)& 4.0 \\[1ex]
6 & 20 (12)  &  8 & -19(9) & -269747(15) & -269743(15) & 3.5\\[1ex]
7 & 17 (13)  &  7 & -19(10) & -273246(16) & -273243(16) & 3.2\\[1ex]
8 & 16 (14)  &  7 & -20(10) & -275583(17) & -275580(17) & 3.0\\[1ex]
9 & 14 (15)  &  7 & -20(10) & -277221(18) & -277218(18) & 2.9\\[1ex]
10 & 13 (16)  &  7 & -20(10) & -278413(19) & -278410(19) & 2.7\\[1ex]
11 & 12 (16)  &  7 & -21(10) & -279308(19) & -279305(19) & 2.6\\[1ex]
12 & 11 (17)  &  7 & -21(10) & -279996(20) & -269993(20) & 2.5\\[1ex]
\hline\hline 
\end{tabular}
\end{center}
\caption{Corrections and results for the difference (1) in the hydrogen and deuterium atoms and
isotopic shift. The one-loop self-energy ($\Delta^{G}(n)$), the vacuum polarization ($\Delta^{VP}(n)$)
contribution and the two-loop correction  $\Delta^{II}(n)$ are the same in both atoms.}
\vskip 1cm
\end{table}

\begin{table}
\begin{center}
\begin{tabular}{||c|c|c|c||}
\hline
\hline
&\multicolumn{3}{|c||}{}\\[-1ex]
 &\multicolumn{3}{|c||}{Correction to $n^3\cdot\left(\Delta E^{I}(ns)-\Delta E^{II}(ns)\right)$ 
(in kHz)}\\[1ex]
\hline
&&&\\[-1ex]
$n$&$(Z\alpha)^6m^2/M$&$(Z\alpha)^4m^3/M^2$& Total\\[1ex]
\hline 
&&&\\[-1ex]
$1$   &  0  &  26  &   26 \\[1ex]
$2$   &  1  &  -13  &  -12  \\[1ex]
$3$   &  2  & -26 &  -24 \\[1ex]
$4$   &  2  & -33  &  -31 \\[1ex]
$5$   &  2  & -36  &  -35 \\[1ex]
$6$   &  2  & -39  &  -37 \\[1ex]
$7$   &  2  & -41  &  -39 \\[1ex]
$8$   &  2  & -42  &  -41 \\[1ex]
$9$   &  2  & -43  &  -42 \\[1ex]
$10$   &  2  & -44  &  -43 \\[1ex]
$11$   &  2  & -45  &  -43 \\[1ex]
$12$   &  2  & -46  &  -44 \\[1ex]
\hline\hline 
\end{tabular}
\end{center}
\caption{Recoil contributions to difference of the definitions.}
\vskip 1cm
\end{table}

\end{document}